\documentclass[showpacs,preprintnumbers,amsmath,amssymb]{revtex4}
\usepackage{bm}
\usepackage{graphicx}
\usepackage{dcolumn}
\newcommand{\be}{\begin{equation}}
\newcommand{\ee}{\end{equation}}
\newcommand{\bea}{\begin{eqnarray}}
\newcommand{\eea}{\end{eqnarray}}
\newcommand{\nn}{\nonumber}
\newcommand{\p}{\phi}
\newcommand{\vp}{\varphi}
\newcommand{\rd}{\partial}

\newcommand {\pb}{\bar \phi}

\begin{document}
\title{$\phi^4$ model on a circle}
\author{F. Loran}
 \email{loran@cc.iut.ac.ir}
 \affiliation{Department of  Physics, Isfahan University of Technology (IUT), Isfahan,  Iran}

 \begin{abstract} The four dimensional critical scalar theory at equilibrium with a thermal bath at temperature $T$
 is considered. The thermal equilibrium state is labeled by $n$ the winding number of the vacua around the compact imaginary-time
 direction which compactification radius is $1/T$. The effective action for zero
 modes  is a three dimensional $\phi^4$ scalar theory in which  the mass of the the scalar field is proportional to
 $n/T$ resembling the Kaluza-Klein dimensional reduction. Similar results are obtained for the theory at zero temperature but in a one-dimensional
 potential well. Since parity is violated by the vacua with
 odd vacuum number $n$, in such cases there is also a cubic term in the effective potential. The  $\phi^3$-term contribution to the vacuum shift at
 one-loop  is of the same order of the contribution from the $\phi^4$-term in terms of the coupling constant of the four dimensional theory but
 becomes negligible as $n$ tends to infinity. Finally, the relation between the scalar classical
 vacua and the corresponding $SU(2)$ instantons on $S^1\times{\mathbb R}^3$ in the 't~Hooft ansatz is studied.
 \end{abstract}
 \maketitle
\section{Introduction}
 In four dimensions, massless $\phi^4$ model with non-positive potential is equivalent to $SU(2)$ Yang-Mills theory
 in the 't~Hooft ansatz \cite{tHooft,Actor,Solitons}. Scalar theories with
 nonpositive potentials are also familiar in gravity theories
 \cite{Maldacena} while they may be problematic in quantum field theory.

 In this paper we obtain solutions of massless $\phi^4$ model with nonpositive potential which are periodic in one direction.
 We show that this periodic solution can be used for
 three purposes. The first one is to study the critical $\p^4$
 theory at finite temperature. The second one is to study the theory
 in a one-dimensional potential well by imposing the Dirichlet
 boundary condition on $\p$ to be vanishing at the boundaries. The
 third application of this solution is to obtain $SU(2)$
 instantons on $S^1\times{\mathbb R}^3$ using the 't~Hooft ansatz.

 The organization of the paper is as follows. In section \ref{D1} after a brief review of scalar field theory at finite temperature, we
 consider the massless $\p^4$-model in one dimension where we study the
 periodic solution of the equation of motion to be used in the
 following sections. In section \ref{D4} we use the result of
 section \ref{D1} to study the four dimensional model at finite
 temperature. The resulting three dimensional effective action for
 zero modes is shown to be a massive $\p^4$ theory. The mass of the
 scalar field is proportional to the winding number of the classical
 vacua, considered as the thermal equilibrium state, around the compact imaginary-time
 direction. In section \ref{Dir} we study the massless
 $\p^4$ theory at zero temperature in a potential well. In this case
 the effective theory is realized in two different sectors corresponding to even
 and odd vacuum numbers. For even vacuum numbers, parity is
 conserved and the interaction is given by a $\p^4$
 term. In the case of odd vacuum number, parity is violated and
 a $\p^3$ interaction is added. We study and compare the contribution form both interaction terms to the vacuum
 shift at one-loop and verify that for large vacuum numbers, the
 $\p^3$-term contribution is negligible in comparison with the
 contribution from the $\p^4$ term. In section \ref{YM} we briefly discuss the $SU(2)$ instantons corresponding
 to the periodic solutions in the 't~Hooft ansatz. There we show that indeed the Yang-Mills field equation corresponds to the critical theory
 with potential $V(\p)\sim -e^2\p^4$  where $e$ is the gauge field coupling constant. Section \ref{con}
 is devoted to conclusion and is closed by discussing the dependence of the entropy of the thermal vacua
 on the corresponding winding number.
\section{$D=1$ $\phi^4$-model at thermal equilibrium}\label{D1}
 The static properties of finite temperature QFT can be derived from the
 partition function ${\cal Z}=\mbox{tr} e^{-H/T}$ where $H$ is the
 Hamiltonian of the quantum field theory and $T$ is the temperature.
 For a simple theory with boson fields $\p$ and Euclidean action
 $S(\p)$ the partition function is given by the functional
 integral
 \be
 {\cal Z}=\int[d\p]\exp\left[-S(\p)\right],
 \label{D1-0-0}
 \ee
 where $S(\p)$ is the integral of the Lagrangian density
 ${\cal L}(\p)$,
 \be
 S(\p)=\int_0^{1/T}dt \int d^dx {\cal L}(\p),
 \label{D1-0-1}
 \ee
 and the field $\p$ satisfies periodic boundary conditions in the
 imaginary-time direction,
 \be
 \p(t=0,x)=\p(t=1/T,x).
 \ee
 The equation of motion of the $\phi^4$ model in one
 dimension, defined by the Euclidean action,
 \be
 S=\int dt \left(\frac{1}{2}\p'^2-\frac{g}{4}\p^4\right).
 \label{D1-0}
 \ee
 where a $'$ denotes one time derivation with respect to $t$, is the following non-linear Laplace
 equation,
 \be
 \phi''+g\phi^3=0,
 \label{D1-1}
 \ee
 This equation can be easily integrated once to obtain,
 \be
 \frac{1}{2}\phi'^2+\frac{g}{4}\phi^4=c.
 \label{D1-2}
 \ee
 For $c=0$ the solution $\phi\sim t^{-1}$ is singular at $t=0$.
 For $c>0$, defining $c=L^{-4}$ one obtains,\footnote{In \cite{ref1} similar wave functions are obtained for a massive harmonic oscillator.}
 \be
 \phi=\frac{1}{L}\left(\frac{4}{g}\right)^{1/4}\mbox{sn}\left(\left.g^{1/4}\frac{t}{L}\right|-1\right),
 \label{D1-3}
 \ee
 in which $\mbox{sn}(u|m)=\sin(\vp)$ is the Jacobi elliptic function  in which
 $\vp=\mbox{am}(u|m)$ is the inverse of Jacobi elliptic function of
 the first kind,  $F(\vp|m)$ defined by the relation,\footnote{For definition and calculations on elliptic functions one
 can use the software Mathematica5, Wolfram Research, Inc.}
 \be
 F(\vp|m)=\int_0^\vp\left(1-m\sin^2\theta\right)^{-1/2}d\theta.
 \label{D1-4}
 \ee
 Defining,
 \be
 u(\vp)=F(\vp|-1)=\int_0^\vp d\theta\frac{1}{\sqrt{1+\sin^2\theta}},
 \label{D1-5}
 \ee
 one can easily verify that $u(\vp)$ is a periodic function,
 \be
 u(\vp+ 2n\pi)=u(\vp)+4nK(-1),\hspace{1cm}n\in{\mathbb N},
 \label{D1-6}
 \ee
 in which $K(m)=F(\frac{\pi}{2}|m)$ denotes the complete elliptic integral of the first
 kind. Consequently the solution (\ref{D1-3}) is periodic,
 \be
 \p(t)=\p\left(t+n\frac{4LK(-1)}{g^{1/4}}\right),\hspace{1cm}n\in{\mathbb N}.
 \label{D1-7}
 \ee
 By identifying the period with $1/T$, one can determine $L$ in
 terms of $T$, as follows,
 \be
 L_n=\frac{g^{1/4}}{4K(-1)}\frac{T^{-1}}{n},\hspace{1cm}n\in{\mathbb N}.
 \label{D1-8}
 \ee
 Thus, there exist a set of {\em classical vacua}
 $\p_n$ with winding number $n\in{\mathbb N}$ around the compact imaginary-time
 direction given by,
 \be
 \p_n(t)=\sqrt{\frac{2}{g}}\omega_n\mbox{sn}(\omega_nt|-1),\hspace{1cm}
 \omega_n=n\omega_1,
  \label{D1-9}
 \ee
 where $\omega_1=4K(-1)T$. It is straightforward to calculate the action corresponding to
 $\p_n$, which is given by
 \be
 S(\p_n)=n^4\frac{\left(4K(-1)\right)^4}{3g}T^3.
 \label{D1-10}
 \ee
 Consequently there is  {\em an action barrier},
 \be
 \Delta S\sim \frac{\left(4K(-1)\right)^4}{3g}T^3,
 \label{D1-11}
 \ee
 separating different vacuum states.
 Given
 the vacua $\phi_n$, it is natural to search for the corresponding kink solutions interpolating between different vacua.
 We leave this question as an open problem. In section \ref{YM} we
 obtain $SU(2)$  instantons on $S^1\times{\mathbb R}^3$ in the 't~Hooft ansatz corresponding
 to $\phi_n$.
 \section{$D=4$ $\phi^4$-model at thermal equilibrium}\label{D4}
 In this section we study the $D=4$ $\phi^4$-model at thermal
 equilibrium corresponding to the vacua $\p_n$ given by
 Eq.(\ref{D1-9}). The idea is similar to the Kaluza-Klein dimensional reduction (see e.g. \cite{KK}).
 Considering a $d+1$-dimensional spacetime with one
 compactified dimension ${\cal M}=\mathbb{R}^{1,d-1}\times S^1$, a general (scalar) field $\sigma(x^\mu,t)$ in which $x^\mu$'s are coordinates of
 $\mathbb{R}^{1,d-1}$ and $t$ is the coordinate on $S^1$, can be decomposed via a Fourier transformation into its zero mode $\sigma_0(x^\mu)$ and
 the Kaluza-Klein modes $\sigma_i(x^\mu)$
 which correspond to the $i$-th winding state along $S^1$. In the original Kaluza-Klein method,
 the classical vacuum state is unique and corresponds
 to  $\sigma=0$. If the there are different local vacuum states in the theory, it is natural to
 anticipate the emergence of  different effective theories for zero-modes characterizing the corresponding vacuum state. This is the
 case that one encounters when zero-modes
 of the $D=4$ $\phi^4$-model at thermal equilibrium corresponding to the vacua $\phi_n$ is
 considered. In this theory, at any local vacua $\phi_n$, the emergence of a Kaluza-Klein mode gives a transition to a different vacua by classically
 penetrating through the barrier (\ref{D1-11}). The classical transition rates can be obtained
 by calculating the interaction term $g\int\phi^4$ in which $\phi$ is replaced with the corresponding Kaluza-Klein mode
 expansion. These terms are negligible at high temperature/weak coupling limit as can be verified from Eq.(\ref{D1-11}).
 Of~course, similar to the Kaluza-Klein dimensional
 reduction method, in order to reconstruct the original theory in the decompactification
 limit $T\to 0$, all of these terms should be considered.
 In the following, we do not consider the Kaluza-Klein excitations and restrict ourselves to
 the effective  action for zero-modes in a local vacua labeled by $\phi_n$.

 Assuming that,
 \be
 \p(\vec x,t)=\p_n(t)+T^{1/2}\pb(\vec x),
 \label{D4-1}
 \ee
 we obtain the three dimensional critical action for zero-mode
 $\p(\vec x)$. The coefficient $T^{1/2}$ is considered in the definition
 of $\p(\vec x)$ to insure that its classical mass-dimension is equal to $1/2$, the
 mass dimension of free scalar fields in three dimensions. One
 should note that since we are considering the critical scalar
 theory at finite temperature, the only mass scale at hand is the
 background temperature $T$ to adjust the mass dimension in the
 dimensional reduction procedure.

 The effective action can be obtained simply by inserting
 Eq.(\ref{D4-1}) into the action,
 \be
 S[\p]=\int_0^{1/T}dt\int d^3x
 \left(\frac{1}{2}\rd_\mu\p\rd^\mu\p-\frac{g}{4}\p^4\right).
 \label{D4-2}
 \ee
 In the formal expansion of $S[\p]=S[\p_n+T^{1/2}\pb]$,
 \be
 S[\p]=S[\p_n]+T^{1/2}\int\pb\left(\frac{\delta S}{\delta \p}\right)_{\p_n}+S_n[\pb],
 \label{D4-3}
 \ee
 the linear term in $\pb$ is vanishing since $\p_n$ is a solution of the equation of motion. Using Eq.(\ref{D1-10}) one verifies that here,
 \be
 \frac{S[\p_n]}{V}=n^4\frac{\left(2K(-1)\right)^4}{3g}T^3,
 \label{D4-4}
 \ee
 where $V=\int d^3 x$ is the volume of the {\em room}. The
 effective action is given by $S[\pb]$,
 \be
 S_n[\pb]=\int
 d^3x\left(\frac{1}{2}\sum_{i=1}^3(\rd_i\pb)^2-\frac{1}{2}m_n^2\pb^2-g_3\pb^3-\frac{g_4}{4}\pb^4\right),
 \label{D4-5}
 \ee
 in which $g_4=T g$ and,
 \bea
 \label{D4-6}
 m_n^2&=&3gT\int_0^{1/T}\p_n^2=6n\omega_n\left(4E(-1)-4K(-1)\right),\\
 \label{D4-7}
 g_3&=&gT^{3/2}\int_0^{1/T}\p_n=0,
 \eea
 where $E(m)$ gives the complete elliptic integral,
 \be
 E(m)=\int_0^{\pi/2}\sqrt{1-m\sin^2\theta}d\theta.
 \ee
  $K(-1)\simeq 1.31$ and $E(-1)\simeq 1.91$. Consequently, the effective action for zero
 modes is a three dimensional $\phi^4$ scalar theory in which the mass of the scalar field are given
 by
 \be
 m_n=nm_1,\hspace{1cm}n\in {\mathbb N},
 \label{D4-11}
 \ee
 resembling the Kaluza-Klein reduction,
 where
 \be
 m_1=\left\{6(4K(-1))\left[4E(-1)-4K(-1)\right]\right\}^{1/2}T\simeq
 8.6 T.
 \label{D4-12}
 \ee
 \section{Dirichlet boundary condition}\label{Dir}
 In this section we study the $D=4$ $\p^4$ theory at zero temperature with Dirichlet boundary
 condition,
 \be
 \p(0,\vec x)=\p(\ell,\vec x)=0,
 \label{dir1}
 \ee
 instead of imposing the periodicity condition $\p(t)=\p(t+1/T)$.

 Rewriting the identity (\ref{D1-6}) as,
 \be
 u(\vp+ n\pi)=u(\vp)+2nK(-1),\hspace{1cm}n\in{\mathbb N},
 \label{dir2}
 \ee
 and noting that $\mbox{sn}(0)=0$, one easily verifies that the Dirichlet boundary condition $\p(\ell)=0$ is satisfied
 by assuming that
 \be
 \ell=\frac{2nLK(-1)}{g^{1/4}}, \hspace{1cm}n\in{\mathbb N},
 \label{dir3}
 \ee
 from which one instead of Eq.(\ref{D1-8}) obtains,
 \be
 L_n=\frac{g^{1/4}}{2K(-1)}\frac{\ell}{n},\hspace{1cm}n\in{\mathbb N}.
 \label{dir4}
 \ee
 Thus, here the set of {\em classical vacua}
 $\p_n$ is given by Eq.(\ref{D1-9}) in which one should assume that
 \be
 \omega_1=2K(-1)T.
 \label{dir5}
 \ee
 All the results of section \ref{D4} are still valid
 after replacing $1/T\to\ell$ and $\omega_n\to \omega_n/2$ or replacing $4K(-1)$ and
 $4E(-1)$ by $2K(-1)$ and $2E(-1)$ respectively.

\begin{figure}
\includegraphics{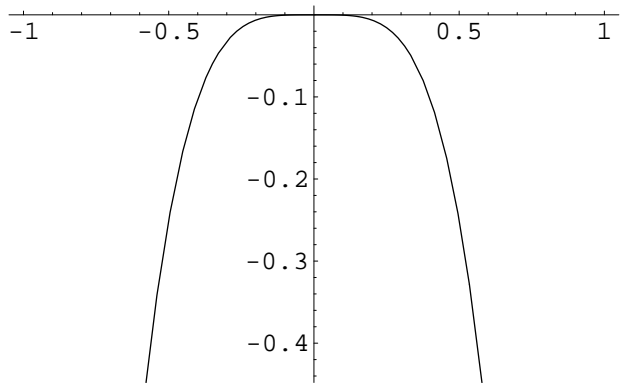}
\caption{\label{even} The effective potential: the even-vacua .}
\end{figure}

\begin{figure}
\includegraphics{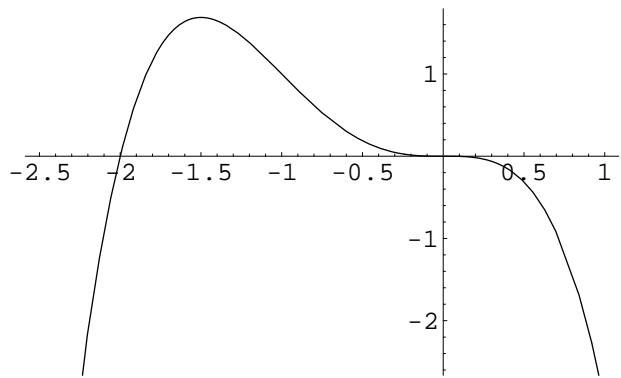}
\caption{\label{odd} The effective potential: the odd-vacua.}
\end{figure}

 Since parity is violated in the four dimensional theory by the vacua $\p_n$ with
 odd vacuum number $n$, it is natural that the parity violating term $g_3\pb^3$ now appear in the expansion (\ref{D4-3}) in this
 case,
 \be
 g_3=\frac{g}{\ell^{3/2}}\int_0^{\ell}\p_n^2=\frac{\pi}{\ell^{3/2}}\sqrt{\frac{g}{2}}\Delta(n),
 \label{dir6}
 \ee
 where
 \be
 \Delta(n)=\left\{\begin{array}{lll}1,&&\mbox{$n$ odd}\\
 0,&&\mbox{$n$ even}\end{array}\right.
 \label{dir7}
 \ee
 The general form of the effective potential for equilibrium
 states given by vacua with even and odd vacuum number $n$ are
 plotted in Fig.~\ref{even} and Fig.~\ref{odd} respectively.
 From these figures one verifies that the $\p^3$-term do not affect
 the theory dramatically for example by generating a local minima. In the following we compute the one-loop contribution to the scalar
 field self-energy for the even and odd cases and show that for large $n$, the vacuum shift in two sectors are similar.
 To this aim, we assume the validity of perturbation around the
 zero of the potential at $\p=0$. As we will show at the end of this
 section even in the case of the odd-vacua, where the
 potential is of the form plotted in Fig.~\ref{odd}, one should do
 perturbation around $\p=0$.

 The one-loop contribution to the self-energy from the $\p^3$ term,
 depends on $p$ the momenta of the external line,
 \bea
 \Sigma^{(3)}_n(p)&=&\frac{\Delta(n)\pi^2}{2}\frac{g}{\ell^3}\int d^3k\frac{1}{k^2+m_n^2}\frac{1}{(\vec k+\vec
 p)^2+m_n^2}\nn\\
 &=&\Delta(n)\pi^4\frac{g}{\ell^3}\frac{2}{p}\sin^{-1}\left(\frac{1}{\sqrt{1+4m_n^2/p^2}}\right).
 \label{1l-1}
 \eea
 From this result, the one loop contribution from the $\p^3$ term
 to the vacuum to vacuum transition (the {\em unobservable} vacuum shift)
 can be obtained,
 \be
 \Sigma^{(3)}_n(0)=\Delta(n)\pi^4\frac{g}{\ell^3}\frac{1}{m_n} \sim
 \frac{\Delta(n)}{n}\frac{g}{\ell^2}.
 \label{1l-2}
 \ee
 The one-loop contribution to the self-energy from the $\p^4$ term,
 is divergent. The divergent term is linear in the  cut-off $\Lambda$ and
 is given by the one-loop diagram for mass-less scalar ($n=0$).
 \bea
 \Sigma^{(4)}_n&=&-\frac{g}{4\ell}\int d^3 k\frac{1}{k^2+m_n^2}\nn\\
 &=&\Sigma^{(4)}_0+\frac{g}{\ell}\pi^2m_n,
 \label{1l-3}
 \eea
 where $\Sigma^{(4)}_0$ is the one-loop vacuum shift for the $n=0$ case,
 \be
 \Sigma^{(4)}_0=-\frac{g}{4\ell}\int d^3 k\frac{1}{k^2}=-\frac{g}{\ell}\pi \Lambda.
 \label{1l-4}
 \ee
 Consequently,
 \be
 \Sigma^{(4)}_n-\Sigma^{(4)}_0=\frac{g}{\ell}\pi^2m_n\sim n\frac{g}{\ell^2}.
 \label{1l-5}
 \ee
 The $\p^4$ theory in three dimensions is super renormalizable.
 After renormalization the vacuum shift is given by,
 \bea
 \Sigma_n&=&\Sigma^{(3)}_n+\Sigma^{(4)}_n-\Sigma^{(4)}_0\nn\\
 &=&\frac{g}{\ell}\pi^2m_n\left(1+\frac{\Delta(n)\pi^2}{m_n^2\ell^2}\right).
 \label{1l-6}
 \eea
 Consequently the $\phi^3$ term slightly modifies the vacuum shift,
 but, the relative difference between the vacuum shift for the even and
 odd cases, decreases as $1/n^2$. This result is
 consistent with our physical intuition because for large $n$ the global behavior of the potential barrier is important thus Fig.~\ref{even} and
 Fig.~\ref{odd} seem more similar to each other as $n$ increases.

 The only thing that we should check is whether the perturbation
 point $\p=0$ we used for both the even and odd cases is the correct
 perturbation base point in the odd case. This can be verified as
 follows. The potential term plus the mass term in the odd-case is
 given by,
 \be
 \tilde V_n(\p)=\frac{1}{2}m_n^2\p^2+g_3\p^3+\frac{g_4}{4}\p^4.
 \label{1l-7}
 \ee
 If there is another point $\p=\p_0$ around which we could do the
 perturbation, then $\tilde V(\p-\p_0)$ should, first of all, be
 of the form,
 \be
 \tilde V(\delta\p)=\mbox{const.}+\frac{1}{2}M_0^2(\delta\p)^2+{\cal
 O}(\delta\p)^3,
 \label{1l-8}
 \ee
 for some constant $M^2_0$. This means that $\p_0$ should be a stationary point of $\tilde
 V(\p)$. To obtain the stationary point, we have to solve the
 equation,
 \be
 \delta\tilde V(\p)/\delta\p=m_n^2\p+3g_3\p^2+g_4\p^3=0.
 \label{1l-9}
 \ee
 this equation in addition to $\p=0$ has another solution if
 \be
 9g_3^2>4m_n^2g_4.
 \label{1l-10}
 \ee
 inserting $g_4=g/\ell$, $m_n\simeq 4.3n/\ell$ and $g_3$ from Eq.(\ref{dir6}) into Eq.(\ref{1l-10}) one verifies that a solution
 like $\p_0$ exists if
 \be
 n^2<0.6.
 \label{1l-11}
 \ee
 Since the $\p^3$ term exists only for odd $n$, i.e. $n\ge 1$, therefore the only solution to Eq.(\ref{1l-9})
 is $\p=0$ and consequently the one-loop calculations above are
 correct.
 \section{$SU(2)$ instanton on $S^1\times{\mathbb R}^3$}\label{YM}
  In this section we discuss $SU(2)$ insantons corresponding to
  the vacua  $\p_n$, in the 't~Hooft ansatz \cite{tHooft}. The 't~Hooft ansatz for the Yang-Mills potential $A^a_\mu$
  is given by,
 \be
 A^a_\mu=\eta^a_{\mu\nu}\rd^\nu\psi/\psi,
 \label{ym1}
 \ee
 where $\eta^a_{\mu\nu}$ are the 't~Hooft tensors,
 \be
 \eta^a_{\mu\nu}=\epsilon_{0a\mu\nu}+i\eta_{a\mu}\eta_{0\nu}-i\eta_{a\nu}\eta_{0\mu},
 \label{ym2}
 \ee
 in which $\eta_{\mu\nu}=(-,+,+,+)$ is the Minkowski metric.
 Assuming that $\psi=\psi(x^0)$, the gauge field $A^a_\mu$ and the
 fieldstrength $F_{\mu\nu}$ are given as follows,
 \be
 A^a_\mu=i\eta_{a\mu}\p,
 \label{ym3-1}
 \ee
 and
 \be
  F^a_{ij}=-e\epsilon_{aij}\p^2,\hspace{1cm}  F^a_{0i}=i\delta_{ai}\rd_0\p,
 \label{ym3-2}
 \ee
 in which $\p=\rd_t\psi/\psi$ and $e$ is the gauge field coupling constant.
 The field equation $\rd^\mu F^a_{\mu\nu}+e\epsilon^{abc}A^{b\mu}
 F^c_{\mu\nu}=0$, reads,
 \be
 \rd_0^2\p-2e^2\p^3=0.
 \label{ym4}
 \ee
 Thus the field equation for an instanton is given by Eq.(\ref{ym4})
 after a Wick rotation $x^0\to ix^0$,
 \be
 \rd_t^2\p+g\p^3=0,
 \label{ym5}
 \ee
 where $g=2e^2$. Using Eq.(\ref{ym3-2}) the instanton action,
 \be
 S_{\mbox{YM}}=-\frac{1}{4}\int d^4x F^a_{\mu\nu}F^{a\mu\nu},
 \label{ym6}
 \ee
 satisfies the identity $S_{\mbox{YM}}=3S[\p]$ where $S[\p]$ is the
 action of the critical scalar theory in $D=4$ given in Eq.(\ref{D4-4}).

 Consequently, the set of solutions $\p_n$ for the critical scalar
 theory gives a set of solution to the $SU(2)$ instanton field
 equation. Furthermore the 't~Hooft ansatz provide a powerful
 motivation to study the critical scalar theory with potential
 $V(\p)\sim -\p^4$  since as is observed above, the coupling
 constant $g\sim e^2$ in Eq.(\ref{ym5}) can not be negative.
 \section{Conclusion}\label{con}
 The massless $\p^4$ model in one dimension has a set of periodic
 solutions,
 \be
 \p_n(t)=\sqrt{\frac{2}{g}}\omega_n\mbox{sn}(\omega_nt|-1),\hspace{1cm}
 \omega_n=n\omega_1,
 \label{con1}
 \ee
 if the scalar potential is $V(\p)\sim-\p^4$. These
 solutions can be used to study the theory  at
 finite temperature in higher dimensions or to study such theories
 in  a one dimensional potential well with Dirichlet boundary
 condition on the scalar field to be vanishing on the boundaries.
 The resulting effective action is a scalar theory in one dimension
 lower. Here we studied dimensional reduction $d=4 \to d=3$. The
 mass of scalar field in the effective theory appeared to be
 proportional to $n$, resembling the Kaluza-Klein dimensional reduction. In the thermal theory, $n$ is the winding number of the classical vacua around
 the compact time direction but in the theory with Dirichlet boundary condition it denotes  the vacuum
 number.
 In the thermal theory, the scalar interaction in the effective theory is given by a $\p^4$
 term. The effective action in the case of $\p^4$ model in one
 dimensional potential well with Dirichlet boundary condition, should be studied in two different
 sectors. If the vacuum number is even, the effective interaction is
 given by a $\p^4$ term. But if the vacuum number is odd, due to parity violation, there is also a cubic term in the effective
 potential. Both theories of course are super renormalizable.

 Finally one can use $\p_n$ to construct $SU(2)$ invariant solutions
 of the $SU(2)$ Yang-Mills field equation on $S^1\times{\mathbb R}^3$, in the 't~Hooft  ansatz,
 \be
 {A^{(n)}}^{a}_\mu=i\eta_{a\mu}\p_n\hspace{1cm}n\in{\mathbb N}.
 \label{con2}
 \ee
 Considering $A^a_\mu$ in Eq.(\ref{con2}) as an instanton the corresponding
 action can be shown to be equivalent to,
 \be
 S_{\mbox{YM}}=n^4\frac{\left(4K(-1)\right)^4}{g\ell^3}.
 \label{con3}
 \ee
 where $\ell$ is the radius of the $S^1$.

 Our motivation for the present work has been to generalize the
 classical Fubini's solution \cite{Fubini} of the massless phi-fourth model. The
 Fubini's solution is invariant under the de Sitter subgroup of the
 full conformal symmetry group of the classical massless phi-forth
 model. His motivation has been to find a natural mass-scale in the
 physics of hadrons. In \cite{ds} we showed that the Fubini's vacua, in the phi-fourth model with non-positive
 potential, can be interpreted as an open FRW de~Sitter background. Furthermore, semiclassical arguments showed that the
 entropy associated to the Fubini's vacua is equivalent to the entropy of a de~Sitter
 vacua. This is a good sign for the relevance of non-positive potentials to physics specially when opposed to
 the phi-fourth model with a
 positive potential which is known to be trivial, see e.g. \cite{Jackiw}. The connection between
 the Fubini's solution and the dS vacua is recently studied in the
 context of M-theory in \cite{Seb}.

 The Fubini's solution is invariant under none of the translations of the
 conformal symmetry group.
 Thus an interesting question is whether there exist a classical vacuum invariant under some translations
 if not all \cite{ds}. As we saw above, such solutions at least provide new instantons of the SU(2) Yang-Mills theory
 which are probably  useful in a braneworld scenario. In such a scenario and probably in the physics
 of superconductors, the scalar theory itself is interesting as it is exhibiting a new mass generating mechanism.

 We close this section by giving some comments on the dependence of
 the entropy ${\cal S}_n$ of vacua $\p_n$ on $n$, the winding number, in the
 case of the scalar theory at thermal equilibrium. Classically, one
 may define an entropy by,
 \be
 T \delta{\cal S}=\delta S,
 \label{con4}
 \ee
 Since the action $S_n$ is not a linear function of $n$, this
 formula will be applicable only for large values of $n$. In this limit, $\delta n^4\simeq 4n^3$, thus equation (\ref{con4}) can be integrated once
 to obtain,
 \be
 {\cal S}_n=n^4\frac{\left(4K(-1)\right)^4}{g}T^2,\hspace{1cm}n\gg1.
 \label{con5}
 \ee
 To obtain the entropy semi-classically, one might determine the
 quantum state $\left|\psi_n\right>$ corresponding to the classical vacua $\p_n$ and define the entropy by the
 relation,
 \be
 \left<\psi_n|\psi_n\right>\sim e^{{-\cal S}_n}.
 \label{con6}
 \ee
 To construct the quantum state $\left|\psi_n\right>$, one may
 proceed as follows. One counts the number of plane waves $e^{ikx}$ with a
 given momenta $k$ superposed to construct $\p_n$ and then create
 the same number of free-particle states from $\left|0\right>$, the
 state of nothing. The resulting spectrum of free-particle states describes the quantum state $\left|\psi_n\right>$ corresponding to the
 classical vacua $\p_n$. Consequently,
 \be
 \left|\psi_n\right>\sim\exp\left({\sum_m
 \sqrt{A_m^{(n)}}a^\dag_m}\right)\left|0\right>,
 \label{con7}
 \ee
 where $A_m^{(n)}$ are given by the Fourier transform of $\p_n$,
 \be
 \p_n(t)=\sum_mA_m^{(n)}\sin(2\pi mTt).
 \label{con8}
 \ee
 Thus to obtain $A^{(n)}_m$ one should calculate the following
 integral,
 \be
 A^{(n)}_m=\frac{\omega_n}{\pi}\sqrt{\frac{2}{g}}\int_0^{2\pi}\mbox{sn}\left.\left(n\frac{K(-1)}{\pi}\theta\right|-1\right)\sin(m\theta)d\theta.
 \label{con9}
 \ee
 We expect that the entropy obtained in this way become equivalent
 to the result of Eq.(\ref{con4}) for large $n$.
 \section*{Acknowledgement}
 The financial support of Isfahan University of Technology (IUT) is
 acknowledged.



\begin{thebibliography}{99}
 \bibitem{tHooft} G. 't Hooft, Phys.Rev.Lett. 37, (1976) 8;\\ G. 't
  Hooft, Phys.Rev. D 14, (1976) 3432.
 \bibitem{Actor}A. Actor, Rev. Mod. Phys. 51, (1979) 461.
 \bibitem{Solitons} F. Loran, Phys. Rev. D 71, 126003 (2005), hep-th/0501189.
 \bibitem{Maldacena} J. Maldacena and  C. Nunez, Int.J.Mod.Phys. A16 (2001) 822-855, hep-th/0007018.
 \bibitem{ref1}C. A. A. de Carvalho, R. M. Cavalcanti, E. S. Fraga, and S. E. Joras, Annals of Physics (n.Y.) 273,  (1999)
 146, {\em ibid}  Phys. Rev. E 61 (2001) 6392.
 \bibitem{KK}  L. O'Raifeartaigh, {\em The Dawning of Gauge Theory},
 Princeton Series in Physics, Princeton University Press, 1997.
 \bibitem{Fubini}S. Fubini, Nuovo Cim.A34, (1976) 521.
 \bibitem{ds} F. Loran, {\em Fubini vacua as a classical de Sitter
 vacua}, hep-th/0612089, to appear in Mod.Phys.Lett. A.
 \bibitem{Jackiw}R. Jackiw, Carlos Nunez and  S.-Y. Pi, Phys. Lett. A347 (2005)
 47.
\bibitem{Seb} S. de Haro and A. C. Petkou, {\em Instantons and Conformal
Holography}, hep-th/0606276'\\
S. de Haro, I. Papadimitriou and A. C. Petkou, {\em Conformally
Coupled Scalars, Instantons and Vacuum Instability in AdS4},
hep-th/0611315.
 \end{thebibliography}
\end{document}